\documentclass{llncs}

\usepackage[english]{babel}
\usepackage{url}
\usepackage{amsmath}
\usepackage{amssymb}
\usepackage{graphicx}
\usepackage[pdf]{pstricks}
\usepackage{pst-tree}
\usepackage{subfigure}

\renewcommand{\bar}{\neg}

\newcommand{\compl}{\bar}

\newcommand{\clauset}{F}

\newcommand{\mfalse}{\textup{\textbf{f}}}
\newcommand{\mtrue}{\textup{\textbf{t}}}
\newcommand{\ta}{\varphi}

\usepackage[bookmarks=false]{hyperref}

\title{Concurrent Cube-and-Conquer\thanks{The second author is supported by DARPA contract number N66001-10-2-4087. The third author is supported by FWF, NFN Grant S11408-N23 (RiSE).}}
\author{Peter van der Tak\inst{1} \and Marijn J.H. Heule\inst{1,2} \and Armin Biere\inst{3}}
\institute{
Delft University of Technology, The Netherlands \and
University of Texas at Austin, United States \and
Johannes Kepler University Linz, Austria
}

\date{\today}

\def\CONST#1{\mbox{\tt{#1}}}

\def\S#1{\mbox{\emph{#1}}}
\def\K#1{\textbf{#1}}
\def\N{\\[0pt]}
\def\I{\hspace{2em}}
\def\L#1{\raise .2ex\hbox{\small\tt #1}&}

\newcommand{\phidec}{\varphi_{\mathrm{dec}}}
\newcommand{\ldec}{l_{\mathrm{dec}}}
\newcommand{\phiimp}{\varphi_{\mathrm{imp}}}

\newcommand{\Qdec}{\S{Q}_\mathrm{decision}}
\newcommand{\Qres}{\S{Q}_\mathrm{solved}}
\newcommand{\Sid}{\S{S}_\mathrm{id}}

\begin{document}

\maketitle

\begin{abstract}
Recent work introduced the cube-and-conquer technique to solve hard SAT instances. It partitions the search space into cubes using a lookahead solver. Each cube is tackled by a conflict-driven clause learning (CDCL) solver. Crucial for strong performance is the cutoff heuristic that decides when to switch from lookahead to CDCL. Yet, this offline heuristic is far from ideal.
In this paper, we present a novel hybrid solver that applies the cube and conquer steps simultaneously. A lookahead and a CDCL solver work together on each cube, while communication is restricted to synchronization. Our concurrent cube-and-conquer solver can solve many instances faster than pure lookahead, pure CDCL and offline cube-and-conquer, and can abort early in favor of a pure CDCL search if an instance is not suitable for cube-and-conquer techniques.
\end{abstract}

\section{Introduction}
Current satisfiability solvers that target industrial instances are almost always based on the conflict-driven clause learning (CDCL)~\cite{cdcl} technique. A technique with fast heuristics and data
structures that can successfully solve very large instances by propagating decisions and learning additional information when conflicts arise. Yet on small, hard problems lookahead solvers~\cite{lookahead}
perform better by applying much more reasoning in each search node and then recursively splitting the search space until a solution is found.

Recent work~\cite{cube-and-conquer} has shown that the two techniques can be combined successfully, resulting in better performance particularly for very hard instances. The key insight is that lookahead
solvers can be used to partition the search space into subproblems that are easy for a CDCL solver to solve. By first partitioning (\emph{cube}) and then solving each subproblem (\emph{conquer}), some instances can be solved within hours rather than days.
This cube-and-conquer approach, particularly the \emph{conquer} part, is also easy to parallelize.

The challenge to make this technique work lies in developing good heuristics to determine when to stop partitioning and start solving. The current heuristics already give good results, but are far from optimal and require some fine tuning to work well with instances of different difficulty. For example, applying too much partitioning can undesirably increase the run time of otherwise easy instances.

The most important problem in developing a better heuristic is that in the partitioning phase no information is available about how well the CDCL solver will perform on a subproblem. The heuristic is required to estimate this performance, but this is not always reliable. In this work we use an online approach that runs both phases concurrently, and that thereby avoids this problem. We focus less on the parallelization of the \emph{conquer} phase.

Other than improving the performance of cube-and-conquer by replacing this heuristic, the online approach aims to solve another problem: for some instances cube-and-conquer performs worse than CDCL regardless of the configuration of the solvers and heuristics. Our approach is able to quickly identify these instances, in which case the problem can be solved using a classical CDCL search.

We believe that CCC is particularly interesting as part of a portfolio solver, where our predictor can be used to predict whether to apply cube-and-conquer techniques. The authors of SATzilla specifically mention in their conclusion that identifying solvers that are only competitive for certain kinds of instances still has the potential to further improve SATzilla's performance substantially~\cite{satzilla}.


\section{Preliminaries}
\label{sec:preliminaries}

For a Boolean variable $x$,
there are two \emph{literals}, the positive literal, denoted by $x$,
and the negative literal, denoted by $\bar x$.
A \emph{clause}  is a disjunction of literals,
and a {\em CNF formula} is a conjunction of clauses.
A clause can be seen as a finite set of
literals, and a CNF formula as a finite set of clauses.
A \emph{unit clause} contains exactly one literal.
A truth assignment for a CNF formula $\clauset$
is a  function $\ta$ that maps variables in $F$ to
$\{\mtrue,\mfalse\}$.
If $\ta(x) = v$,
then \mbox{$\ta(\compl{x}) = \neg v$},
where $\neg \mtrue = \mfalse$ and $\neg \mfalse = \mtrue$.
A clause $C$ is satisfied by $\ta$ if  $\ta(l) = \mtrue$ for some $l\in C$.
An assignment $\ta$
satisfies $\clauset$ if it satisfies every
clause in $\clauset$.
A \emph{cube}  is a conjunction of literals
and a {\em DNF formula} a disjunction of cubes.
A cube can be seen as a finite set of
literals and a DNF formula as a finite set of cubes.
If $c = (l_1 \land \dots \land l_k)$ is a cube, then $\bar c = (\bar l_1 \lor \dots \lor \bar l_k)$ is a clause.
A truth assignment $\varphi$ can be seen as the cube of literals $l$ for which $\varphi(l) = \mtrue$.
A cube $c$ is satisfied by $\ta$ if $\ta(l) = \mtrue$ for all $l \in c$. An assignment $\ta$ satisfies DNF formula $D$ if it satisfies some cube in $D$.
A DNF formula $D$ is called a {\em tautology} if every full assignment $\ta$ satisfies $D$.
%

\subsection{Cube-and-conquer}
The technique proposed in this work is based on
\emph{cube-and-conquer}~(CC)~\cite{cube-and-conquer}. 
CC was designed for solving very hard instances by partitioning the search space into \emph{cubes} using a lookahead solver (march\_cc), and then solving each cube using an incremental CDCL solver (iLingeling). The key observation made by the authors is that CDCL solvers often solve these cubes very fast, and as a result the two-phase solver is faster than either solver on its own. Additionally, it is natural to parallelize by solving multiple cubes in parallel. In this work we mainly use MiniSAT~2.2~\cite{minisat} as CDCL solver instead of Lingeling, since it is easier to extend.

Cube-and-conquer modifies the lookahead solver to cut off its search based on a \emph{cutoff heuristic}. When the heuristic triggers, the conjunction of decision literals (cube) is stored and the solver continues as if the branch was unsatisfiable. When finished, all cubes are solved incrementally by a CDCL solver, by adding a cube's literals as assumptions~\cite{assumptions} to the original formula and running the search. The disjunction of cubes is a tautology, so that solving each cube individually is equivalent to finding a solution to the original formula.

The cutoff heuristic multiplies the number of assigned decision variables and the total number of assigned variables as an indication of the complexity of the current cube. If this number exceeds a threshold value, the branch is cut off. The threshold is chosen dynamically: it is decreased when lookahead proves unsatisfiability for a branch, because CDCL would likely have solved it faster, and it is also decreased when lookahead descends too deep in the decision tree, which would result in too many cubes. A more detailed explanation follows in Sec.~\ref{sec:cutoff}.

\section{Motivation}

Cube-and-conquer shows strong performance on several hard application benchmarks~\cite{cube-and-conquer}, beating both the lookahead
and CDCL solvers that were used for the cube and conquer steps. However, on many other instances, either lookahead or CDCL
outperforms CC. 
We observed that for benchmarks for which CC has relatively weak performance, two important assumptions regarding the foundations 
of CC do not hold in general. 

First, in order for CC to perform well,  lookahead heuristics must be able to split the search space into cubes that, combined, take less time for CDCL to solve. 
Otherwise, cube-and-conquer techniques are ineffective and CDCL would be the preferred solving technique.
Second, lookahead must be able to refute cubes that are easy for CDCL to solve, and it should not refute cubes that are still hard for CDCL. When this assumption fails, the cutoff heuristic will perform badly, and the cube phase either generates too few cubes and leaves a potential performance gain unused, or generates too many cubes because cubes with fewer decisions are also easy for CDCL to solve.
In this section, we discuss the involved heuristics in more detail, and we discuss how to predict when these heuristics are ineffective. 

In related work on portfolio SAT solving~\cite{satzilla,argosmart,3s} machine learning techniques are used for selection (including parameters) and scheduling
of SAT solvers.  
These techniques are based on measuring several features of instances, which can be characterized as either being static, such as number of
variables and clauses, or dynamic, such as the number of propagated assignments at certain decision depths~(local search or DPLL probing~\cite{satzilla}).  In
this section, we describe a new dynamic feature, which allows us to predict the effectiveness of lookahead and cutoff heuristics in the context of CC and
extensions.


\subsection{Lookahead heuristics}

To compare the performance of CDCL and CC, we ran both solver types\footnote{MiniSAT 2.2 for CDCL; MiniSAT 2.2 and march\_cc (cube phase) and iMiniSAT 2.2 (conquer phase) for CC.
All benchmarks were first preprocessed using Lingeling as suggested for CC in~\cite{cube-and-conquer}.  We used the same version of Lingeling as in~\cite{cube-and-conquer}.}
on all application benchmarks of SAT 2009. CDCL was able to solve 57 more benchmarks than CC within the timeout of 900 seconds (171 vs 114).
For some instances, the performance gap was huge (in favor of CDCL), in particular on satisfiable ones. This can be explained as follows. After a decision, the reduced formula might be harder (or at least not easier) than the original one. This may be caused by ineffective lookahead heuristics. In case a decision hardly reduces the search space, the conquer solver could need to solve two similar problems instead of one, thereby raising the computational costs. On satisfiable formulas this negative effect is expected to be larger, since a single wrong decision might bring the solver in a part of the search space without solutions.

The main reason for this negative effect is that the key assumption underlying CC fails. This assumption expects that
lookahead decision heuristics can select for a formula $F$ a decision variable $x$ in such a way that
$F \cup \{x\}$ and $F \cup \{\bar x\}$ are easier to solve separately than $F$ itself. 
It was shown that for several benchmarks this assumption holds~\cite{cube-and-conquer}. 
However, the results above show that for many benchmarks in the SAT 2009 application suite 
this is not the case. For those, one would like to apply pure CDCL instead of CC.





Ineffective lookahead heuristics can be observed as follows.
Given a formula $F$ and a decision variable $x$, lookahead creates two branches $F \cup \{x\}$ and $F \cup \{\bar x\}$. The branch that reduces the formula the most is called the {\em right} branch, or a {\em discrepancy}.
%
In case lookahead heuristics are effective, then with each decision, but especially each discrepancy the formula becomes much simpler. Thus, after only a few discrepancies, lookahead (or CDCL) should be able to refute the branch. A cube that is reached through many discrepancies suggests that the lookahead 
heuristics have not been effective for that branch.


\subsection{The cutoff heuristic}

The cutoff heuristic is crucial for performance of cube-and-conquer.
Cutting off too early wastes a potential performance gain, but cutting off too deep can result in a large number of instances increasing the total run time. Yet with the current heuristic it is often the case that thousands or millions of cubes are solved almost instantly, while one or two remain and take the majority of the run time. This suggests that the heuristic is not able to properly detect which branches are easy and should be cut off.  

In case this behavior is observed, two complementary actions would be preferred. 
On the one hand, for the many cubes that are solved almost instantly, the cutoff should have taken place earlier (at a smaller cube) to reduce the cost of the cube phase of CC.
On the other hand, for the cubes that require lots of computational resources, the cutoff should have been performed later to use lookahead for further partitioning. 
In short, if this happens -- in CC only in the conquer phase -- then the cutoff heuristic should be considered ineffective.

%
%
%
%
%

\subsection{Predicting when to apply cube-and-conquer}
To predict for which benchmarks CC is competitive, we propose \emph{concurrent cube-and-conquer} (CCC) as follows. During the cube phase of CC, run a CDCL solver in parallel which follows the decisions of the cube solver (details are described in Sec.~\ref{sec:CCC}). By running both solvers simultaneously, the cutoff heuristic becomes obsolete, because the CDCL solver naturally determines whether a cube is easy for CDCL\footnote{Still, cutoff heuristics can lead to reduced resource usage and better performance, as described in Sec.~\ref{sec:cutoff}.}. With the cutoff heuristic out, we only need to predict when lookahead heuristics are ineffective. The following two metrics can be used to predict when this is the case.

First, lookahead techniques appear effective if they can solve some cubes faster than CDCL.
While running the lookahead and CDCL solver in parallel, we count the number of times that lookahead is faster than CDCL.
For benchmarks for which this count is increased very slowly, say less than once per second, we observed that CC was generally not an effective solving strategy.

Second,  if the variable heuristics are effective then 
each discrepancy should result in a large reduction of the formula. Hence after a certain number of discrepancies the solver should be able to refute that branch. 
Preliminary experiments suggest that if CCC finds a leaf with 
over 20 discrepancies early in the search-tree, then lookahead variable heuristics should be considered as ineffective.
Lookahead solvers solve the left branch first as it is heuristically most likely to be satisfiable. In contrast, CCC considers the right branch first so that it can quickly detect if a branch with a large number of discrepancies is encountered.


To predict whether an instance is suitable for (C)CC, these metrics are combined as follows. Run CCC and abort it if it enters a branch with more than 20 discrepancies. If after 5 seconds CCC is still running but 10 or fewer cubes were solved by lookahead, also abort the solver. For aborted instances (unpredicted instances), a pure CDCL search is run instead. For instances that were not aborted (predicted instances) CCC is the preferred solving technique and can continue. The same instances usually work well for CC, but they cannot be detected as easily because CDCL is only used in the conquer phase. In fact, we have not been able to come up with a quick CC-based predictor that works well.

\section{Concurrent cube-and-conquer}
\label{sec:CCC}


This section describes the \emph{concurrent cube-and-conquer} (CCC) technique. We first describe CCC$_\infty$, and extend it later by adding a cutoff heuristic like in CC for better resource utilization. CCC$_\infty$ constructs a decision tree via the lookahead solver and simultaneously runs a CDCL solver on the newest node of this decision tree. Whenever the lookahead solver assigns a decision variable, the new literal is sent to the CDCL solver, which adds it as an assumption and restarts. This is repeated recursively until either solver proves unsatisfiability, which means that the cube is refuted and both solvers backtrack. Whereas CC uses a cutoff heuristic to determine which branches are cut off, CCC$_\infty$ cuts branches off implicitly when CDCL proves unsatisfiability before lookahead makes another decision.

Ideally, this approach is implemented within one solver.
However, due to lack of appropriate data structures, current CDCL solvers only apply lookahead and other forms of preprocessing at the top-level, and not under assumptions.
For instance tree-based lookahead~\cite{marcheq} requires access to all binary clauses at all decision levels,
which can only be accessed in a fast manner by either using full occurrence lists or three watches for non-binary clauses.
Both techniques are not easy to combine with data structures currently used in CDCL solvers.

On the other hand, lookahead solvers lack data structures for conflict analysis and learning, which is essential in CDCL solvers for allowing non-chronological backtracking and
for cutting off repeated parts of the search.
CC and CCC can be seen as two different ways of solving this dilemma by running both types of solvers separately, sequentially in CC and concurrently in CCC$_\infty$.

\smallskip

CC showed to be particularly useful if many cubes were generated, which means that CCC needs frequent synchronization.
To keep the synchronization costs small, CCC$_\infty$ uses asynchronous message queues, where both solvers are peers.
This architecture also makes it easy to integrate other solvers in the future.

The solvers in CCC$_\infty$ communicate using two queues: the decision queue $\Qdec$
and the result queue $\Qres$. Whenever the lookahead solver assigns a decision variable, it pushes the tuple $\langle$cube $c_{id}$,
literal $l_{\mathrm{dec}}$, $\S{backtrackLevel}\rangle$ comprising a uniquely allocated $id$, the decision literal, and the number of
previously assigned decision variables ($\S{backtrackLevel}$). When the CDCL solver reads the new decision from the queue, it already
knows all previous decision literals, and only needs to backtrack to the $\S{backtrackLevel}$ and add $l_{\mathrm{dec}}$ as an
assumption to start solving $c_{id}$. The $id$ is used to identify the newly created cube.
%
%
%

If the CDCL solver proves unsatisfiability of a cube before it receives another decision, it pushes the $c_{id}$ of the refuted cube to $\Qres$. The solver then continues with the parent cube, by backtracking to the level where all but the last decision literal were assigned. When the lookahead solver reads the $c_{id}$ from $\Qres$, it backtracks to the level just above this cube's last decision variable and continues its search as if it proved unsatisfiability of the cube by itself.

The CDCL solver proves unsatisfiability of a cube if it encounters a complementary assignment when attempting to assign one of a cube's literals. This is not necessarily the last literal of the cube, so that it may refute not only the cube corresponding to the latest decision read from $\Qdec$, but also one or more of its parent cubes. Therefore, it sends only $c_{id}$ of the smallest cube which it refuted, which implies that the sub cubes are also unsatisfiable.

To keep track of the cubes that are pending to be solved, both solvers keep the trail of decision literals (or assumptions for the CDCL solver) and the $id$s of the cubes up to and including each decision literal (or assumption).
Whenever either solver proves unsatisfiability of the empty cube, or when it finds a satisfying assignment, the other solver is aborted.

\smallskip

It is possible that the lookahead solver already proved unsatisfiability of a cube when it receives the same result from the CDCL solver. The id is used to discard results on $\Qres$ for cubes that have already been closed. Similarly, it is possible that the lookahead solver makes a decision even though the CDCL solver already proved unsatisfiability of a parent of that cube. In that case the CDCL solver can discard the obsolete item on $\Qdec$.

\subsection{Example}

\begin{figure}
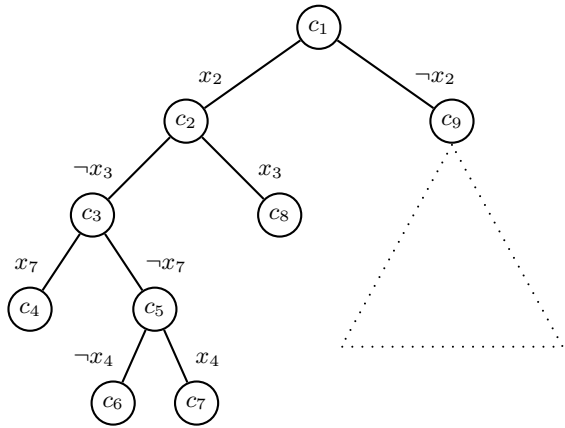

\centering
$
\pstree[levelsep=1.25cm,treesep=.5cm,treefit=loose]{\Tcircle{c_1}}{
	\pstree{\Tcircle{c_2}^{x_2}}{
		\pstree{\Tcircle{c_3}^{\lnot x_3}}{
			\Tcircle{c_4}^{x_7}
			\pstree{\Tcircle{c_5}_{\lnot x_7}}{
				\Tcircle{c_6}^{\lnot x_4}
				\Tcircle{c_7}_{x_4}
			}
		}
		\Tcircle{c_8}_{x_3}
	}
	\pstree[levelsep=3cm]{\Tcircle{c_9}_{\lnot x_2}}{
		\Tfan[linestyle=dotted,fansize=3cm]{}
	}
}
$
\caption{Example decision tree.}
\label{fig:example-tree}
\end{figure}

Consider the decision tree in Figure~\ref{fig:example-tree}. The decisions made by the lookahead solver are displayed on the edges, and each node contains the $c_{id}$ of the cube corresponding to the literals on the path from the root of the tree up to that node. The
$id$'s are incremented based on depth first search.

Assume that $c_4$ has been refuted previously, and both solvers are currently solving $c_6$. Now, if the CDCL solver finds a conflict when assigning assumption $\lnot x_3$, it knows that $c_3$ is unsatisfiable and pushes $c_3$ to $\Qres$. It then removes assumptions $\lnot x_4$, $\lnot x_7$, and $\lnot x_3$, continues with $c_2$, and propagates $x_3$ because it analyzed the conflict and learned something like $(\lnot x_2 \lor x_3)$.

When the lookahead solver reads $c_3$ from  $\Qres$, it will abort its search in $c_6$, skip over $c_7$ and also abort $c_5$, and $c_3$. It continues solving $c_8$ by making decision $x_3$ from  $c_2$. Note that it is possible that when the lookahead solver reads $c_3$ from  $\Qres$, it has already progressed and is solving  $c_7$, or even $c_8$ or $c_9$. When solving $c_7$, the same action can be taken: abort $c_7$, $c_5$, and $c_3$. In case it is at $c_8$ or beyond, then $c_3$ will no longer be part of the trail and the message is skipped because $c_3$ is already known to be unsatisfiable.

Now consider what would happen if the lookahead solver proves unsatisfiability of $c_6$: without sending anything to the CDCL solver, it would backtrack to $c_5$ and then enter the right branch, pushing $(c_7, x_4, 3)$ onto $\Qdec$. If the CDCL solver has not yet solved $c_5$ by the time it reads from $\Qdec$, it backtracks to level 3 ($c_5$), decides $x_4$, and thereby starts solving $c_7$.

\subsection{Implementation}

The listing in Fig.~\ref{fig:march_ccc} shows pseudocode for the implementation of the lookahead solver in CCC$_\infty$. The function is called recursively for each cube that is entered, and all arguments except for the formula $\S{F}$ are initially empty lists. Line 1 allocates a new, unique $\S{id}$ for the cube. Line 2 checks if the CDCL solver has proved any new cubes unsatisfiable. If it did, and if that cube is part of the cube that is currently being solved, the search for this cube is aborted on line 3. If it is not a parent of the current cube, then this result is no longer relevant and is removed from the queue via line 4.

If the cube was not yet solved by CDCL, line 5 adds its id to the id trail $\Sid$, which is the list of all nodes on the path from the root of the decision tree to the current cube. Line 7 sends the id and new decision literal to the CDCL solver, as well as the level at which the literal was added, because the CDCL solver should backtrack to this level before adding the literal. Line 6 only handles the special case of the root cube, which has no decision literal and does not need to be sent to the CDCL solver.
Lines 8 to 13 describe the core of any typical lookahead solver, and are responsible for propagating unit clauses, checking if a solution is found, and otherwise making a new decision and calling $\S{LA\_search}$ recursively for the two cubes that result by adding the literal or its complement to $\phidec$.

\smallskip

Fig.~\ref{fig:minisat_ccc} lists the pseudocode for CCC$_\infty$'s CDCL solver. $\S{S}$ contains the ids of cubes that are currently being solved, much like $\Sid$ in the lookahead solver, and is initially empty.
Lines 4 to 9 handle decision literals sent by the lookahead solver. If a new decision is available on queue $\Qdec$, then the solver pushes the id of the new cube on $\S{S}$, backtracks to the indicated level, and adds the new decision literal on lines 7, 8, and 9 respectively. Line 6 handles a special case that can occur when the lookahead solver makes a decision while the CDCL solver already proved its parent cube unsatisfiable. In that case, $\S{backtrackLevel}$ (the size of the parent cube) will be larger than the number of decisions in the current cube, $|\S{S}|$, which indicates that the decision is no longer relevant and can be ignored. If no decisions are waiting on the queue, line 11 makes a decision using the CDCL solver's heuristics.
Line 16 will detect if the CDCL solver finds a conflict while assigning one of the literals in the cube, in which case lines 17 and 18 will notify the lookahead solver. Since one conflict can prove multiple cubes in $\S{S}$ unsatisfiable, line 17 removes the larger cubes so that line 18 only sends the smallest cube that was proved unsatisfiable. The lookahead solver only needs the smallest cube, because this will implicitly abort the larger cubes too.

The remaining lines are like any other CDCL solver. Restarts are not part of the pseudocode, but should be implemented by backtracking to level $|\S{S}|$ instead of level 0 for CCC$_\infty$ to work correctly. In addition, we reset the restart strategy and reduce the clause database size every time a cube is refuted for better performance. Cubes are solved in the same order as they were generated, and two threads of a parallel solver never solve the same cube (the multijob strategy~\cite{cube-and-conquer}).

\begin{figure}
\vspace{-10pt}
\centering
\begin{tabular}{r@{~~}l}
\L{} \K{global} $\Qdec = \emptyset, \Qres = \emptyset, \S{id} = 0$ \N
\N
\L{}$\S{LA\_search}(F, \phidec, \phiimp, \Sid)$ \N
\L{1}\I $\S{id} := \S{id} + 1$ \N
\L{2}\I \K{while} \K{not} $\Qres.\S{empty}()$ \K{do} \N
\L{3}\I\I \K{if} $\Qres.\S{head}() \in \Sid$ \K{then} \K{return} $\CONST{UNSAT}$ \N
\L{4}\I\I $\Qres.\S{remove}()$ \N
\L{5}\I $\Sid.\S{push}(\S{id})$ \N
\L{6}\I \K{if} $|\phidec| > 0$ \K{then} \N
\L{7}\I\I $\Qdec.add(\langle \S{id}, |\phidec|-1, \phidec.\S{last}() \rangle)$ \N
\L{8}\I $\langle F, \phiimp \rangle := \S{simplify\_and\_learn}(F, \phidec, \phiimp)$ \N
\L{9}\I \K{if} $\phidec \cup \phiimp$ falsify a clause in $F$ \K{then} \K{return} $\CONST{UNSAT}$ \N
\L{10}\I \K{if} $\phidec \cup \phiimp$ assign all variables in $F$ \K{then} \K{return} $\CONST{SAT}$ \N
\L{11}\I $\ldec :=$ $\S{decide}(F, \phidec, \phiimp)$ \N
\L{12}\I \K{if} $\S{LA\_search}(F, \phidec \cup \{\ldec\}, \phiimp, \Sid) = \CONST{SAT}$ \K{then} \K{return} $\CONST{SAT}$ \N
\L{13}\I \K{return} $\S{LA\_search}(F, \phidec \cup \{\lnot\ldec\}, \phiimp, \Sid)$ \N
\end{tabular}
\caption{Pseudocode listing of march\_ccc.}
\label{fig:march_ccc}
\vspace{-10pt}
\end{figure}

\begin{figure}
\centering
\vspace{-10pt}
\begin{tabular}{r@{~~}l}
\L{}        $\S{CDCL\_search}(F)$ \N
\L{1}\I          $\S{S} := \emptyset$ \N
\L{2}\I          \K{forever} \K{do} \N
\L{3}\I\I            \K{if} $\varphi$ assigns all variables in $\S{F}$ \K{then} \K{return} $\CONST{SAT}$ \N
\L{4}\I\I            \K{if} $\Qdec \neq \emptyset$ \K{then} \N
\L{5}\I\I\I              $\langle \S{id}, \S{backtrackLevel}, \ldec \rangle := \Qdec.\S{remove}()$ \N
\L{6}\I\I\I              \K{if} $\S{backtrackLevel} > |\S{S}|$ \K{then} \K{continue} \N
\L{7}\I\I\I              $\S{S}.\S{push}(\S{id})$ \N
\L{8}\I\I\I              $\varphi := \S{backtrack}(\varphi, \S{backtrackLevel})$ \N
\L{9}\I\I\I              $\varphi := \varphi \cup \{\ldec = \mtrue\}$ \N
\L{10}\I\I            \K{else} \N
\L{11}\I\I\I              $\varphi := \varphi \cup \{\S{pickDecisionLiteral}() = \mtrue\}$ \N
\L{12}\I\I            $\varphi := \S{propagate}(\S{F}, \varphi)$ \N
\L{13}\I\I            \K{while} $\varphi$ falsifies a clause in $\S{F}$ \K{do} \N
\L{14}\I\I\I              $\langle \S{conflict}, \S{backtrackLevel} \rangle := \S{analyze}(\S{F}, \varphi)$ \N
\L{15}\I\I\I              \K{if} $\S{conflict} = \emptyset$ \K{then} \K{return} $\CONST{UNSAT}$ \N
\L{16}\I\I\I              \K{if} $\S{backtrackLevel} < |\S{S}|$ \K{then} \N
\L{17}\I\I\I\I              \K{while} $|\S{S}| > \S{backtrackLevel} + 1$ \K{do} $\S{S}.pop()$ \N
\L{18}\I\I\I\I              $\Qres.add(\S{S}.\S{pop}())$ \N
\L{19}\I\I\I            $\S{F} := \S{F} \land \S{conflict}$ \N
\L{20}\I\I\I            $\varphi := \S{backtrack}(\varphi, \S{backtrackLevel})$ \N
\L{21}\I\I\I            $\varphi := \S{propagate}(\S{F}, \varphi)$ \N
\end{tabular}
\caption{Pseudocode listing of minisat\_ccc.}
\label{fig:minisat_ccc}
\vspace{-10pt}
\end{figure}

\subsection{Reintroducing the cutoff heuristic}
\label{sec:cutoff}
One advantage of CC was that the conquer phase could be parallelized efficiently by using multiple CDCL solvers in parallel, each solving
a single cube. With CCC$_\infty$ this is no longer possible, since the lookahead solver will continue with a single branch until it is
solved by either CDCL or lookahead. Additionally, CCC$_\infty$ always uses twice as much CPU time as wall clock time, because the
lookahead and CDCL solvers run in parallel.

To reduce this wasted resource utilization and allow for parallelization of the CDCL solver, we reintroduce the conquer phase by applying a suitable cutoff heuristic. As with CC, we pass cubes from the cube phase to the conquer phase via the file system using the iCNF\footnote{\url{http://users.ics.tkk.fi/swiering/icnf}} format, which is basically a concatenation of the original formula $F$ and the generated cubes as assumptions. An incremental SAT solver iterates over each cube $c_{id}$ in the file, and solves $F \land c_{id}$ until a solution is found or all cubes have been refuted. We use iMiniSAT and iLingeling with four CDCL solvers for the serial respectively parallelized conquer phase, denoted CCC$_\mathrm{mini}$ and CCC$_\mathrm{lgl4}$. We use CCC to refer to the cube phase regardless of what conquer solver is used.

The cutoff heuristic of CC is based on a rough prediction of the performance of CDCL on a cube. Given a cube $c_{id}$, it computes its difficulty\footnote{CC's heuristic has been improved slightly since it was initially published \cite{cube-and-conquer}; it now uses $|\phidec|^2$ instead of $|\phidec|$.}\footnote{The notation is ours.} $d(c_{id}) := |\phidec|^2 \cdot (|\phidec| + |\phiimp|) / n$, where $|\phidec|$ and $|\phiimp|$ are the number of decision and implied variables respectively, and $n$ is the total number of free variables. If $d(c_{id})$ is high, the CDCL solver is expected to solve $c_{id}$ fast.

The cutoff heuristic in CC focuses on identifying cubes that are easy for CDCL to solve. It cuts off a branch if $d(c_{id})$ exceeds a dynamic threshold value $t_{cc}$. Initially $t_{cc} = 1000$, and it is multiplied by 0.7 whenever lookahead solves a cube (because it assumes that CDCL would have solved this cube faster) or when the number of decisions becomes too high (to avoid generating too many cubes). It is incremented by 5\% at every decision to avoid the value from dropping too low.

For CCC, the same heuristic does not work because easy cubes are solved quickly by the CDCL solver. This makes the threshold very unstable so that it quickly converges to 0 or infinity depending on the instance. We therefore use a different heuristic, but using the same difficulty metric $d(c_{id})$.

Easy cubes can be detected better by CCC than by CC, because CCC can detect for which cubes CDCL finds a solution before the lookahead solver does. CCC would ideally cut off these cubes so that they can be solved in parallel. The contrary goes for when the lookahead solver solves a cube: it then seems that lookahead contributes to the search, which means that it is not desirable to cut off.

CCC uses the same difficulty metric $d(c_{id})$ as CC, but a different heuristic for determining the threshold value $t_{ccc}$. If a cube $c_{id}$ is solved by CDCL, the value is updated towards $s := 0.4 \cdot d(c_{id})$, whereas it is updated towards $s := 3 \cdot d(c_{id})$ if $c_{id}$ was solved by lookahead. To avoid too sudden changes, $t_{ccc}$ is not changed to $s$ directly but is filtered by $t'_{ccc} := 0.4 \cdot s + 0.6 \cdot t_{ccc}$. To furthermore avoid the threshold from dropping too low, it is incremented for every cube that is cut off.

\begin{figure}[bt]
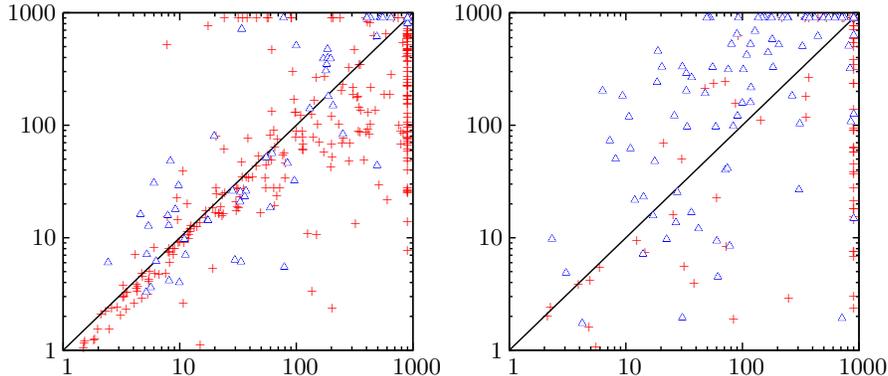

\centering
\vspace{-10pt}
\label{fig:scatter}
\begin{minipage}{0.45\textwidth}
\include{scatter-app}
\end{minipage}
\hfil
\begin{minipage}{0.45\textwidth}
\include{scatter-craft}
\end{minipage}
\vspace{-10pt}
\caption{Scatter plots showing the effect of the performance in seconds of MiniSAT (vertical axis) and CCC$_{\mathrm{mini}}$ (horizontal axis) on benchmarks selected  (${\blue \triangle}$) and not selected (${\red +}$) by the predictor. Left application benchmarks of SAT 2009 and SAT 2011, right crafted instances of SAT 2009 and SAT 2011. Above the line CCC$_{\mathrm{mini}}$ is stronger, below the line MiniSAT is stronger.}
\label{fig:scatter}
\vspace{-10pt}
\end{figure}

\section{Empirical results}
In this section we discuss the performance of the CCC solvers and the effectiveness prediction. We have first run CCC$_\infty$ for 5 seconds on all instances from the application and crafted categories of the SAT 2009 and 2011 competition\footnote{\url{http://www.satcompetition.org/}} and selected only instances where CCC$_\infty$ is not aborted in favor of a pure CDCL search by the predictor. These instances are referred to as \emph{predicted} instances. Since the prediction takes at most 5 seconds and usually much less, we consider the overhead hardly significant. We therefore focus our experiments on predicted instances.\footnote{The sources of the used software and the list of predicted instances are available on \url{http://fmv.jku.at/cccreview}.}

The predictor selects 44 out of 292 instances from the SAT 2009 application suite, and 41 out of 300 from the SAT 2011 application suite. For crafted instances it selects a larger fraction: 70 out of 281 and 99 out of 276 for the 2009 and 2011 crafted suites respectively. As seen in Fig.~\ref{fig:scatter}, the predictor mostly selects instances for which CCC$_\mathrm{mini}$ works well compared to MiniSAT, and there are almost no instances where CCC$_\mathrm{mini}$ times out ($>$900 seconds) and MiniSAT does not. For unpredicted instances combined from both application categories, CCC$_\mathrm{mini}$ solves only 208 instances within a 900 second timeout versus 274 by MiniSAT. For the crafted instances that is 115 for CCC$_\mathrm{mini}$ versus 141 for MiniSAT. We therefore argue that the predictor is very well suited to select the instances where cube-and-conquer works well.

We ran each predicted instance on the following solvers: (C)CC$_\mathrm{mini}$, (C)CC$_\mathrm{lgl4}$, CCC$_\infty$, reference solvers MiniSAT~2.2, March\_rw~\cite{marchrw}, and Lingeling, and parallel solver Plingeling4 (Plingeling with four threads). The CCC solvers all use MiniSAT~2.2 and March\_cc (March\_rw with cube support) concurrently in the cube phase as described in Sec.\ref{sec:CCC}, and the CC solvers only use March\_cc in the cube phase. CCC$_\mathrm{mini}$ and CC$_\mathrm{mini}$ use iMiniSAT in the conquer phase, and (C)CC$_\mathrm{lgl4}$ uses iLingeling with four parallel CDCL solvers in the conquer phase. Before passing an instance to any solver, the instance was preprocessed with Lingeling's -s option.

We report on wall clock time unless stated otherwise. For CCC$_\infty$ the CPU time is twice the wall clock time since two solvers run concurrently. For CCC$_\mathrm{mini}$ the cube phase is
usually short and the run time is dominated by the conquer phase, hence the wall clock and CPU time are often similar. For CCC${}_{\mathrm{lgl4}}$ the times deviate most, as the conquer phase is parallelized efficiently.

\begin{figure}[!b]
\vspace{-10pt}
\include{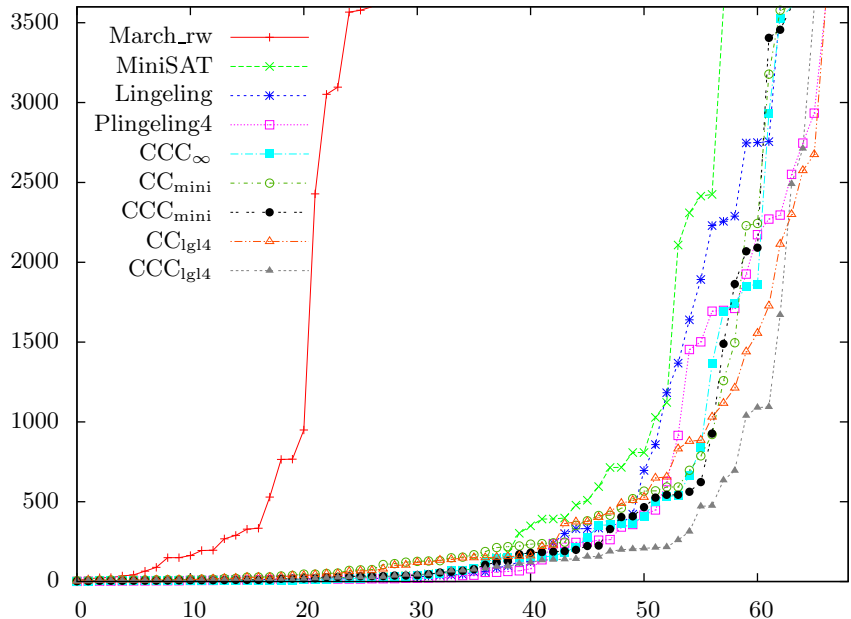}
\vspace{-25pt}
\caption{Cactus plot of various solvers on the application benchmarks of the SAT09 and SAT11 competitions selected by the CCC predictor.}
\label{fig:cactus-app}
\vspace{-10pt}
\end{figure}

\begin{figure}[ht]
\vspace{-15pt}
\include{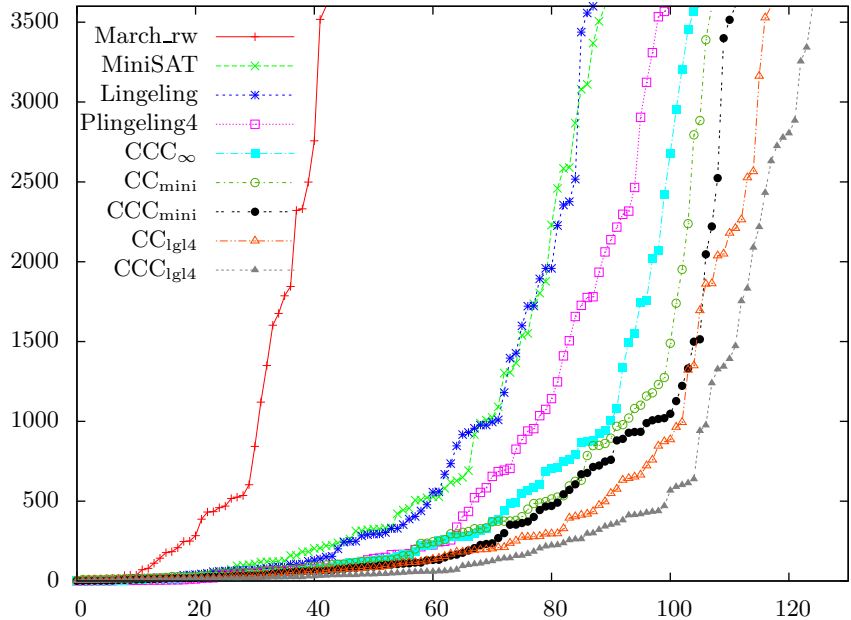}
\vspace{-25pt}
\caption{Cactus plot of various solvers on the crafted benchmarks  of the SAT09 and SAT11 competitions selected by the CCC predictor.}
\label{fig:cactus-craft}
\vspace{-10pt}
\end{figure}

The cactus plots in Fig.~\ref{fig:cactus-app} and Fig.~\ref{fig:cactus-craft} show that all cube-and-conquer techniques are strong on the predicted instances: all (C)CC solvers outperform the three reference solvers in the crafted categories, and perform slightly better on application instances especially with lower timeout values. CCC$_\mathrm{mini}$ solves 3 more instances than MiniSAT within a 3600 second time limit for both application suites: 32 vs 35 and 25 vs 28 for the SAT 2009 and 2011 application instances respectively. The performance on crafted instances is even better: CCC$_\mathrm{mini}$ solves 5 more instances than MiniSAT (52 vs 57) in the SAT 2009 crafted category, and 17 more in the SAT 2011 crafted category (38 vs 55).

The results for CCC$_\mathrm{lgl4}$ show that the cubes generated by CCC can be parallelized well, even though CCC with a single Lingeling solver (not plotted for clarity) performs worse than CCC$_\mathrm{mini}$. For application instances, the differences between Plingeling4 (the winner of the SAT 2011 competition's application category in wall-clock time), CC$_\mathrm{lgl4}$, and CCC$_\mathrm{lgl4}$ are not so large: CCC$_\mathrm{lgl4}$ performs slightly better for lower time limits, but Plingeling4 and CC$_\mathrm{lgl4}$ eventually solve one more instance. For crafted instances, CCC$_\mathrm{lgl4}$ performs best, and solves 7 more than its predecessor CC$_\mathrm{lgl4}$ and 24 more than Plingeling4.

It is interesting to see that CCC$_\infty$ still performs reasonably well, even though it is a very extreme version of CCC where the cube and conquer phases are fully merged. Although it is not the best configuration, it shows that the online usage of CDCL really contributes to the lookahead search: by cutting off leafs early using MiniSAT, CCC$_\infty$ solves many more instances than pure March does. Detailed results show that the wall clock time of CCC$_\infty$ is often slightly higher than CCC$_\mathrm{mini}$, but the biggest problem is that some instances that are solved quickly by other solvers and not at all by CCC$_\infty$. It seems that for some instances, CDCL is not fast enough to cut off enough cubes. Additionally the CPU time is much larger for CCC$_\infty$ because at all times the two solvers run concurrently without idling.

\section{Conclusion}
In this work we proposed an online cube-and-conquer solver that solves the two main limitations of offline cube-and-conquer. First, it is able to predict efficiently on which instances it works well, and abort the search after a few seconds in favor of a pure CDCL solver if not. Second, it does not estimate the performance of CDCL on a cube merely by assuming that it is similar to the performance of lookahead on that cube. This is not true in general so that offline cube-and-conquer is often not able to determine when to stop partitioning and start solving.

The cube-and-conquer solver we proposed runs a lookahead and CDCL solver concurrently to partition the search space. We have seen that this not only implicitly improves the run time of the cube phase, it also allows for better cutoff heuristics so that the generated cubes are easier for a CDCL solver to solve. Like offline cube-and-conquer, our approach allows the conquer phase to be parallelized efficiently.

\bibliography{references}{}
\bibliographystyle{plain}

\end{document}